\documentclass[12pt,a4paper]{article}
\usepackage{amsmath}
\usepackage{graphicx}
\usepackage{amsfonts}
\usepackage{amssymb}
\usepackage{epsfig}
\usepackage{psfrag}
\usepackage{wrapfig}
\usepackage{color}

\definecolor{yellow}{rgb}{1,0.9,0} 
\definecolor{wine}{rgb}{0.5,0,0.4}
\definecolor{lightblue}{rgb}{0.3,0,0.7}
\definecolor{ancientrose}{rgb}{0.7,0,0.4}
\definecolor{cream}{rgb}{1.,1.,0.7}
\definecolor{violet}{rgb}{1.,0.9,0.95}
\definecolor{lightgreen}{rgb}{0.8,1,0.8}
\definecolor{darkgreen}{rgb}{0,0.6,0}
\newcommand{\be}{\begin{equation}}
\newcommand{\ee}{\end{equation}}
\newcommand{\beq}{\begin{eqnarray}}
\newcommand{\eeq}{\end{eqnarray}}

\def\nue{\mathrel{{\nu_e}}}
\def\numu{\mathrel{{\nu_\mu}}}
\def\nutau{\mathrel{{\nu_\tau}}}
\def\nux{\mathrel{{\nu_x}}}

\def\barnue{\mathrel{{\bar \nu}_e}}
\def\barnumu{\mathrel{{\bar \nu}_\mu}}
\def\barnutau{\mathrel{{\bar \nu}_\tau}}
\def\barnux{\mathrel{{\bar \nu}_x}}

\def \lta {\mathrel{\vcenter{\hbox{$<$}\nointerlineskip\hbox{$\sim$}}}}
\def \gta {\mathrel{\vcenter{\hbox{$>$}\nointerlineskip\hbox{$\sim$}}}}

\def\t13{\mathrel{{\sin^2 \theta_{13}}}}
\def\y12{\mathrel{{\tan^2 \theta_{12}}}}

\begin{document}

\begin{titlepage}

\vspace*{0.2cm}

\begin{center}


{\Large \bf Physics of supernova neutrinos: \\
flavor conversion effects} \\


\vspace{0.8cm}
{\large
Cecilia Lunardini$^{\ast}$} \\

\vspace{0.2cm}
{Institute for Advanced Study, Einstein drive, 08540 Princeton,
New Jersey, USA}

\end{center}
\begin{abstract}
I 
review the effects of flavor conversion of neutrinos from
stellar collapse due to masses and mixing, and discuss the motivations
for their study. I consider in detail the sensitivity of certain
observables (characteristics of the energy spectra of $\nu_e$ and
$\bar{\nu}_e$ events) to the 13-mixing ($\sin^2\theta_{13}$) and to
the type of mass hierarchy/ordering (sign[$\Delta m^2_{13}$]). These
observables are: the ratio of average energies of the spectra, $r_E
\equiv \langle E \rangle / \langle \bar{E}\rangle $, the ratio of
widths of the energy distributions, $r_{\Gamma} \equiv \Gamma/
\bar{\Gamma}$, the ratios of total numbers of $\nu_e$ and
$\bar{\nu}_e$ events at low energies, $S$, and in the high energy
tails, $R_{tail}$. I show that regions in the space of observables
$r_E$, $r_{\Gamma}$, $R_{tail}$ exist in which certain mass hierarchy
and intervals of $\sin^2\theta_{13}$ can be identified or
discriminated.  
%
Finally, 
I discuss the potential of studying regeneration effects on
$\nue$ and $\barnue$ in the matter of the Earth and point out that
both the observation or exclusion of these effects 
lead to model-independent information on $\sin^2\theta_{13}$ and the
mass hierarchy.  The extraction of this information would highly
benefit from the presence of a new, large, long lived $\nue$ detector,
and from the progress of theoretical predictions of the fluxes and
energy spectra of the neutrinos originally produced in the star.

\end{abstract}

\vskip10pt
\noindent
{\it PACS:} 14.60.Pq, 97.60.Bw.

\noindent {\it Keywords:} neutrino conversion; matter effects; supernova.

\vfil
\noindent

\footnoterule
{\small
Contribution to the 11th Annual International Conference on Supersymmetry  and the Unification of the Fundamental Interactions (SUSY 2003), Tucson, Arizona, June 5--10 2003.\\
$^{\ast}$ E-mail: lunardi@ias.edu\vskip-1pt\noindent}


%
\thispagestyle{empty}
\end{titlepage}

\setcounter{page}{1}

\section{Introduction and motivations}
\label{sec1}

The mechanism of neutrino flavor conversion due to masses and flavor
mixing has been recently established by the combination of the results
of solar neutrino detectors and those of the KamLand experiment
\cite{Giunti:2003vf}. Results from the detection of atmospheric
neutrinos and the preliminary data from the K2K experiment
\cite{Fogli:2003th} strongly support the existence of this
phenomenon.

From the analysis of all the available data, we get a partial
reconstruction of the neutrino masses $m_i$ (the label $i=1,2,3$ denotes
the neutrino mass eigenstates) and of the mixing matrix $U$, defined
by $\nu_\alpha=\sum_{i} U_{\alpha i} \nu_i$, where $\nu_\alpha$
($\alpha=e,\mu,\tau$) are the flavor eigenstates.  Using the standard
parameterization of the mixing matrix in terms of three angles,
$\theta_{12},\theta_{13},\theta_{23}$, we have:
\begin{equation}
m_2^2 - m_1^2 \equiv   \Delta m^2_{21} = (4  - 30) \cdot 10^{-5}
{\rm eV}^2, ~~~~~~\tan^2 \theta_{12}=0.25 - 0.85 ~,
\label{sunpar}
\end{equation}
from solar neutrinos and KamLand, and
\begin{equation}
m_3^2 - m_2^2 \equiv    \Delta m^2_{32} = \pm (1.5 - 4) \cdot 10^{-3}
{\rm eV}^2, ~~~~~~~ \tan^2 \theta_{23} = 0.48 - 2.1 ~
\label{atmpar}
\end{equation}
from atmospheric neutrinos. The sign of $\Delta m^2_{32}$ is unknown. The two possibilities, $\Delta m^2_{32} \approx \Delta m^2_{31} > 0$ and
$\Delta m^2_{32} \approx \Delta m^2_{31} < 0$, are
referred to as {\it normal}  and {\it inverted} mass hierarchies/ordering respectively  (abbreviated as n.h. and i.h. in the text).

The mixing angle $\theta_{13}$, which describes the $\nu_e$ content
of the third mass eigenstate, $\nu_3$, is still unmeasured. We have an
upper bound from the CHOOZ and Palo Verde experiments \cite{Apollonio:1999ae,Boehm:2000vp}:
\begin{equation}
\sin^2 \theta_{13} \lta 0.02~.
\label{ue3}
\end{equation}
The identification of the neutrino mass hierarchy and the
determination of $\theta_{13}$ have become the main issues of further
studies.
\\

To achieve these, and other important goals, the study of neutrinos
from core collapse supernovae is particularly interesting.  Indeed,
these neutrinos are produced and propagate in {\it unique} physical
conditions of high density and high temperature, and therefore can
manifest effects otherwise unaccessible.  As will be discussed in the
following, due to the very large interval of matter densities realized
there, the interior of a collapsing star is the only environment where
a neutrino of a given energy undergoes two MSW resonances, associated
to the two mass squared splittings of the neutrino spectrum.
This implies a  richer phenomenology of flavor conversion, and
therefore wider possibilities to probe the relevant parameters, with
respect to the case of neutrinos in solar system, where only one
resonance, i.e. one mass splitting, is relevant at a time.

It is important to consider, however, that the study of supernova
neutrinos is not exempt of problems. The main obstacle is the absence
of a ``Standard Model'' for supernova neutrinos, i.e. of precise
predictions for the fluxes and energy spectra of neutrinos of
different flavors originally produced in the star. The features of
these fluxes depend on many details of the neutrino transport inside
the star and, in general, on the type of progenitor star
\cite{Keil:2002in}.

Since observables depend both on the features of the original fluxes
and on the flavor conversion effects, it is clear that 
the extraction of information on the neutrino mixing and on the
neutrino mass spectrum requires a careful consideration of
astrophysical uncertainties.

\section{Properties of supernova neutrino fluxes and density profile of the star}
\label{sec2}
Neutrinos and antineutrinos of all the three flavors are produced in a
supernova and emitted in a burst of $\sim 10$ seconds duration.  At a
given time $t$ from the core collapse the original flux of the
neutrinos of a given flavor, $\nu_\alpha$, can be described by a
``pinched'' Fermi-Dirac (F-D) spectrum\footnote{An alternative
parameterization has been recently suggested in \cite{Keil:2002in}.},
\begin{eqnarray}
F^0_\alpha(E,T_\alpha,\eta_\alpha,L_\alpha,D) =
\frac{L_\alpha}{4\pi D^2 T^4_\alpha F_3(\eta_\alpha)} \frac{E^2}{e^{E/
T_{\alpha}-\eta_\alpha}+1}~,
\label{eq3}
\end{eqnarray}
where $D$ is the distance to the supernova (typically $D\sim 10$ kpc for a
galactic supernova), $E$ is the energy of the neutrinos, $L_\alpha$ is the
luminosity in the flavor $\nu_\alpha$, and $T_\alpha$ represents an
effective temperature. 
The normalization factor equals: $F_3(\eta_\alpha)\equiv \int_0^{\infty} dx~ {x^3}/{(e^{x-\eta_\alpha}+1)} $.
Supernova simulations provide the indicative values of the average energies \cite{Keil:2002in}: 
\be
\langle E_{\bar e} \rangle = (14 - 22)~  {\rm MeV}, ~~~ \langle E_{x} \rangle/\langle E_{\bar e}\rangle = (1.1 - 1.6), ~~~
 \langle E_{e} \rangle/\langle E_{\bar e}\rangle = (0.5 - 0.8),~~~
\label{temp}
\ee and the typical value of the (time integrated) luminosity in each
flavor: $L_\alpha \sim (1 - 5) \cdot 10^{52}~{\rm ergs } $.  The
luminosities of all neutrino species are expected to be approximately
equal, within a factor of two or so \cite{Keil:2002in}. The $\numu$ and
$\nutau$ ($\barnumu$ and $\barnutau$) spectra are equal with good
approximation, and therefore the two species can be treated as a
single one, $\nux$ ($\barnux$).
The pinching parameter $\eta_\alpha$ can vary between $0$ and $\sim 3$ for $\nue$ and $\barnue$, while smaller pinching is expected for $\numu,\nutau$: $\eta_\mu = \eta_\tau \sim 0 - 2$.

The matter density profile met by the neutrinos can be approximated,
at least in the first few seconds of their emission, by that of the
progenitor star \cite{BBB}. The latter is well described by the radial
power law \cite{BBB}:
\begin{eqnarray}
\rho(r)=10^{13}~ C  \left(\frac{10 ~{\rm km}}{r} \right)^3~{\rm g\cdot
cm^{-3}} ~,
\label{eq4}
\end{eqnarray}
with $C\simeq 1 - 15$.

\section{Conversion in the star, jump probability and $\theta_{13}$}
\label{sec3}

\begin{figure}[ht]
\centerline{\epsfxsize=4.9in\epsfbox{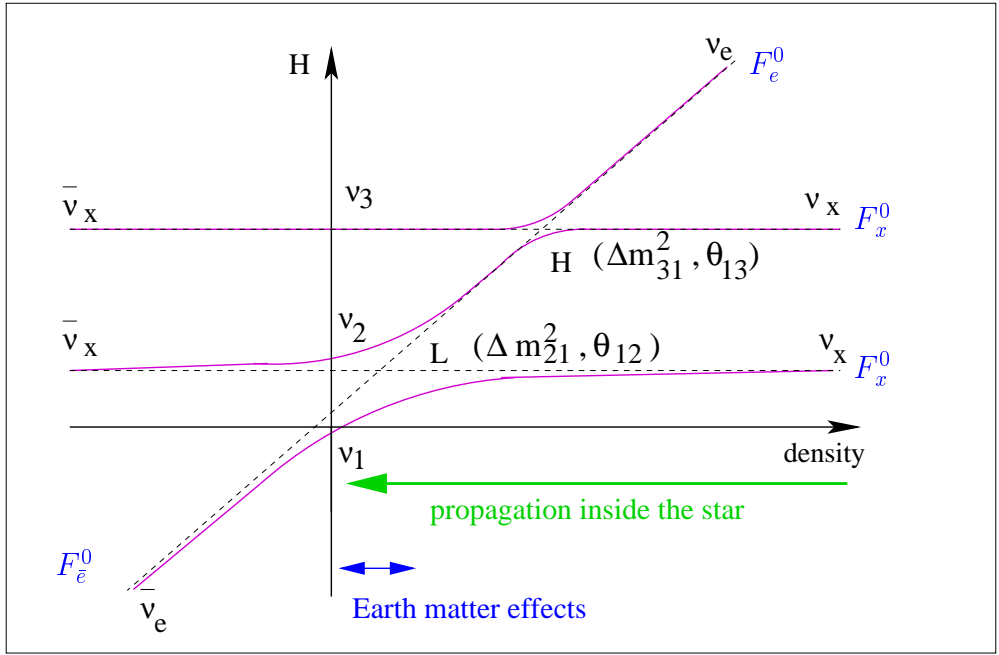}}   
\caption{The level-crossing diagram for normal mass hierarchy. The solid curves represent the eigenvalues of the Hamiltonian in matter. \label{fig:1}}
\end{figure}

Let us consider the conversion of neutrinos as they propagate from the
production region outward in the star, for the case of normal mass
hierarchy ($\Delta m^2_{32}>0$).  As shown in Fig. \ref{fig:1}
(positive density semi-plane), the eigenvalues of the Hamiltonian in
matter and the flavor composition of its eigenstates change with the
variation of the matter density along the neutrino trajectory.

At production, the mixing is suppressed due to the very large density
($\rho \sim 10^{11}~{\rm g\cdot cm^{-3}}$), therefore the eigenstates
of the Hamiltonian coincide with the flavor states.  At lower
densities, the neutrinos undergo two MSW resonances
(level-crossings). The inner resonance (H) is governed by the
parameters $\Delta m^2_{32}$ and $\theta_{13}$ and is realized at
$\rho \sim 10^{3} ~{\rm g\cdot cm^{-3}}(10 {\rm MeV} / E)$.
The probability of transition between the eigenstates of the
Hamiltonian (jump probability) in this resonance, $P_H$, strongly
depends on $\theta_{13}$ as discussed later in this section.  The
second resonance, (L) is determined by $\Delta m^2_{21}$ and
$\theta_{12}$ and happens at lower density, $\rho \sim (30 - 140) (10
{\rm MeV} / E) ~{\rm g\cdot cm^{-3}}$.  For the values of parameters
in Eq. (\ref{sunpar}) the jump probability in this resonance is
negligibly small (adiabatic propagation).  The neutrinos leave the
star as mass eigenstates and therefore do not oscillate on the way
from the star to the Earth.  If they cross the Earth before detection,
oscillations are restarted due to Earth matter
effects (see e.g. \cite{Dighe:1999bi}).

Since they have opposite sign of the matter potential, antineutrinos do not undergo any resonance in the matter of the star  (negative density semi-plane in Fig. \ref{fig:1}).

As an effect of conversion, the $\nue$ and $\barnue$ fluxes in the detector, $F_{ e}$ and $F_{\bar e}$, are  combinations of the original $\nue$ and $\nux$ ($\barnue$ and $\barnux$)  fluxes. Considering for simplicity the case of no Earth crossing, one gets:
\beq
&&F_{ e} = P_H \sin^2 \theta_{12}  F_{ e}^0 + (1- P_H \sin^2 \theta_{12}) F_{ x}^0 ~~,
\nonumber \\
&&F_{\bar e} =
\cos^2 \theta_{12} F_{\bar e}^0 + \sin^2 \theta_{12}  F_{\bar x}^0 ~.
\label{fluxes2}
\eeq

For inverted hierarchy ($\Delta m^2_{32}<0$), the H resonance is in the antineutrino channel, while the L resonance is unaffected.  In this case the fluxes in the detector equal: 
\beq
&&F_{ e} = \sin^2 \theta_{12}  F_{ e}^0 +\cos^2 \theta_{12}  F_{ x}^0 ~~,
\nonumber \\
&&F_{\bar e} =
P_H \cos^2 \theta_{12} F_{\bar e}^0 + (1- P_H \cos^2 \theta_{12})  F_{\bar x}^0 ~.
\label{fluxes2ih}
\eeq
As expected, here the jump probability $P_H$ appears in the expression of the $\barnue$ flux, in contrast with Eqs. (\ref{fluxes2}).
\\

In summary, the supernova neutrino signal is sensitive to the mass
hierarchy and to $\theta_{13}$ for the following reasons: (i)
depending on the hierarchy, the H resonance affects either neutrinos
or antineutrinos; (ii) the observed $\nue$ or $\barnue$ fluxes depend
on the value of $\theta_{13}$ via the jump probability $P_H$. The
latter can be calculated using the Landau-Zener formula and the
profile (\ref{eq4}). The result is: 
\beq &&P_H=\exp \left[-\left(
\frac{ 1.08 \cdot 10^{7} ~{\rm MeV}}{ E }\right)^{2/3} \left(
\frac{\Delta m^2_{32} }{ 10^{-3} ~{\rm eV}^2}\right)^{2/3} C^{1/3}
\sin^2 \theta_{1 3} \right] ~.
\eeq
It follows that three regions exist:

\noindent
(i)  Adiabaticity breaking region: $\t13 \lta 10^{-6} \left({E}/{10 {\rm MeV}}\right)^{2/3}$, where $P_H \simeq 1$;\\
(ii)  Transition region: $\t13 \sim  (10^{-6} - 10^{-4})\cdot \left({E}/{10 {\rm MeV}}\right)^{2/3}$, where $0\lta P_H \lta 1$;\\
(iii)  Adiabatic region: $\t13 \gta 10^{-4} \left({E}/{10 {\rm MeV}}\right)^{2/3}$, where $P_H \simeq 0$.

Notice that if $P_H=1$ (adiabaticity breaking region) 
Eqs. (\ref{fluxes2}) and (\ref{fluxes2ih}) coincide. Thus, we get
equal predictions for normal and inverted hierarchy and any
sensitivity to the mass hierarchy is lost.  Furthermore, from
Eqs. (\ref{fluxes2}) and (\ref{fluxes2ih}) it is easy to see that, in
the extreme case in which the original fluxes in the different flavors
are equal ($F^0_{\bar e}=F^0_{\bar x}$, $F^0_{e}=F^0_{ x}$), conversion
effects cancel and one has $F_e=F^0_e$, $F_{\bar e}=F^0_{\bar e}$.

\section{Earth matter effects}
\label{sec:earth}

If the neutrinos cross the Earth before detection, they undergo
regeneration effects due to the interaction with the matter of the
Earth. Indeed, the matter density in the Earth, $\rho \sim (1 - 13 )
~{\rm g\cdot cm^{-3}}$, is close to the density at the L resonance in
the star (see Sec. \ref{sec3}), for which the mixing angle
$\theta_{12}$ is resonantly enhanced.  This implies that the amplitude
of neutrino oscillations in the Earth can be significant and lead to
observable effects.


The results (\ref{fluxes2}) and (\ref{fluxes2ih}) can be immediately generalized by the replacements:
\be
\sin^2 \theta_{12} \rightarrow P_{2e} \hskip1.5truecm 
\cos^2 \theta_{12} \rightarrow P_{1 \bar e} ~, 
\label{repl}
\ee where $P_{2e}$ ($P_{1 \bar e}$) is the probability that a neutrino
(antineutrino) arriving at Earth in the state $\nu_2$ ($\bar \nu_1$)
is detected as $\nue$ ($\barnue$) in the detector. It depends on the
oscillation parameters $\theta_{12}$ and $\Delta m^2_{21}$, on the
Earth density profile, on the neutrino arrival direction and on the
neutrino energy.  The dependence on the energy has an oscillatory
character, resulting in characteristic distortions of the energy
spectra of observed events.  The distortions are different for
detectors at different locations on the planet.

As an example, consider two detectors, D1 and D2, with D2 shielded by
the Earth and D1 unshielded.  The differences of the $\nue$ and
$\barnue$ fluxes in the two detectors follow from Eqs. (\ref{fluxes2}), (\ref{fluxes2ih}) and (\ref{repl}). For n.h. they are:
\beq
&&F^{D2}_{ e}- F^{D1}_{ e}= P_H (P_{2e}-\sin^2 \theta_{12}) (F_{ e}^0 - F_{ x}^0) ~~,
\nonumber \\
&&F^{D2}_{\bar e} - F^{D1}_{\bar e}=(P_{1\bar e}-\cos^2 \theta_{12}) (F_{\bar e}^0 - F_{\bar x}^0) ~,
\label{fluxesearth}
\eeq
while for i.h. one gets:
\beq
&&F^{D2}_{ e}- F^{D1}_{ e}= (P_{2e}-\sin^2 \theta_{12}) (F_{ e}^0 - F_{ x}^0) ~~,
\nonumber \\
&&F^{D2}_{\bar e} - F^{D1}_{\bar e}=P_H  (P_{1\bar e}-\cos^2 \theta_{12}) (F_{\bar e}^0 - F_{\bar x}^0) ~.
\label{fluxesearth_ih}
\eeq 
The dependence of these differences on the neutrino mass
hierarchy and on $\theta_{13}$ has the same origin as that of
conversion effects in the star and can be described in analogous terms
(Sec. \ref{sec3}).
 
\begin{figure}[ht]
\centerline{\epsfxsize=4.3in\epsfbox{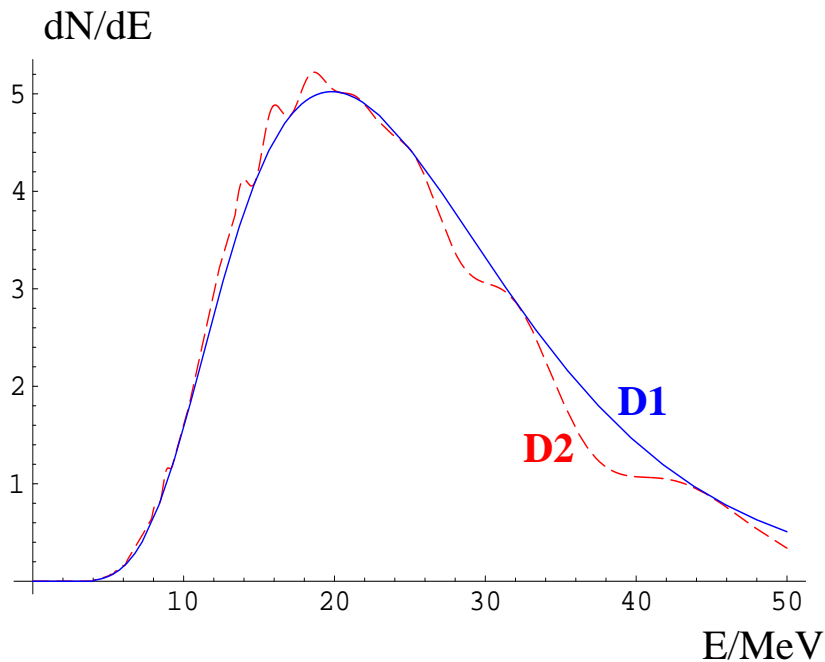}}   
\caption{Predicted energy spectra of positrons (arbitrary units on the
vertical axis) from the $\barnue + p \rightarrow n + e^+$ reaction at
two identical water Cerenkov detectors, one of which (D2) is shielded
by the Earth with nadir angle $\theta_n=70^\circ$. I used the
detection efficiency of the SuperKamiokande detector and the
parameters: $T_{\bar e}=5$ MeV, $T_{\bar x}=7$ MeV, normal hierarchy
(or i.h. with $P_H=1$) and $\Delta m^2_{21}=7\cdot 10^{-5}~{\rm
eV^2}$. Equal luminosities have been taken in $\barnue$ and $\barnux$
and the detailed Earth density profile from ref. \cite{PREM} has been
used. \label{figearth}}
\end{figure}

Figure \ref{figearth} gives an illustration of the energy spectra of
events predicted from Eq. (\ref{fluxesearth}) for two identical water
Cerenkov detectors.

\section{Probing $\theta_{13}$ and the mass hierarchy}
\label{sec4}

There are several approaches to probe the neutrino oscillation parameters
and at the same time take into account the uncertainties on the
features of the original fluxes: \\

\noindent
1.  to perform a global fit of the data, determining both the
oscillation parameters and the parameters of the original fluxes
simultaneously \cite{Barger:2001yx,Minakata:2001cd}. With this method
a completely general analysis is not possible due to the large number
of parameters involved.  \\

\noindent 
2. to single out and study (numerically and analytically)
 specific observables which (i) have maximal
sensitivity to the oscillation parameters of interest and (ii) whose
dependence on the astrophysical uncertainties is minimal or well
understood \cite{Lunardini:2003eh}. \\

\noindent 
3. to study Earth matter effects
\cite{Dighe:1999bi}, \cite{Lunardini:2001pb}--\cite{usprep}.
\\

\noindent 
4. to study the effects of shock-wave propagation on the neutrino
   signal. The shockwave driving the supernova explosion modifies the
   density profile of the star at the resonance points, thus changing
   the conversion pattern inside the star and the observed neutrino
   energy spectra
   \cite{Schirato:2002tg,Takahashi:2002yj,Lunardini:2003eh,Fogli:2003dw}.
   \\

Here I summarize some aspects of the methods 2. \cite{Lunardini:2003eh}
and 3. \cite{usprep}.

\subsection{Observables}
\label{sub41}

Let us consider the method of the observables (item 2.).  A good
prescription to find observables which fit the criteria of sensitivity
and uncertainty-freedom stated above ((i) and (ii)) is to consider the
parameters describing the energy spectra of events induced by $\nue$,
and the same parameters for the $\barnue$-induced spectra, and take
their ratios.

For instance, let us consider the spectra of $\nue$ events at the SNO
detector from the CC scattering on deuterium, $\nue+ d \rightarrow
p+p+e^- $, and the $\barnue$ events at the SuperKamiokande 
detector from inverse beta decay, $\barnue + p \rightarrow n + e^+$.

We can define the following four observables: (1)
the ratio of  the average energies, $r_E$,  and (2) the ratio of the widths, $r_\Gamma$, of the $\nue$ and $\barnue$-induced spectra:
\be
r_E = \frac{\langle E\rangle}{\langle \bar E\rangle}~, \hskip 1.5truecm  r_{\Gamma}  \equiv  \frac{\Gamma}{\bar{\Gamma}}~;
\label{r-e}
\ee 
(3) the ratios of the numbers of $\nue$ and $\barnue$ events in the  low energy tails, $S$, and (4)  in the high energy tails, $R_{tail}$:
\be
S\equiv  \frac{N_e(E<E'_L)}{N_{\bar e}(E<\bar E'_L)}~, \hskip 0.7truecm R_{tail}(E_L,\bar{E}_L)\equiv \frac{N_{e}(E>E_L)}{N_{\bar e}(E>\bar{E}_L)}~.
\label{eq:rat}
\ee
Here the overbarred quantities  refer to antineutrino spectra, and the width $\Gamma$ is defined as $\Gamma \equiv (\langle E^2\rangle/\langle E\rangle^2-1)^{1/2}$. The high and low energy cuts, $E_L,\bar{E}_L,E'_L,\bar E'_L$ can be suitably chosen to optimize the analysis \cite{Lunardini:2003eh}.

\subsection{Distinguishing between extreme possibilities: scatter plots}
\label{sub42}

Let us consider the three extreme cases:\\ 
A. Normal hierarchy with
$P_H=0$, i.e. $\sin^2\theta_{13}\gta 10^{-4}$ (adiabatic region, see
Sec. \ref{sec3});\\ 
B. Inverted hierarchy with $P_H=0$;\\ 
C. $P_H=1$,
corresponding to $\sin^2\theta_{13}\lta 10^{-6}$, with normal or
inverted hierarchy (recall that results do not depend on the hierarchy
in this case, see Sec. \ref{sec3}).

\begin{figure}[ht]
\centerline{\epsfxsize=5.3in\epsfbox{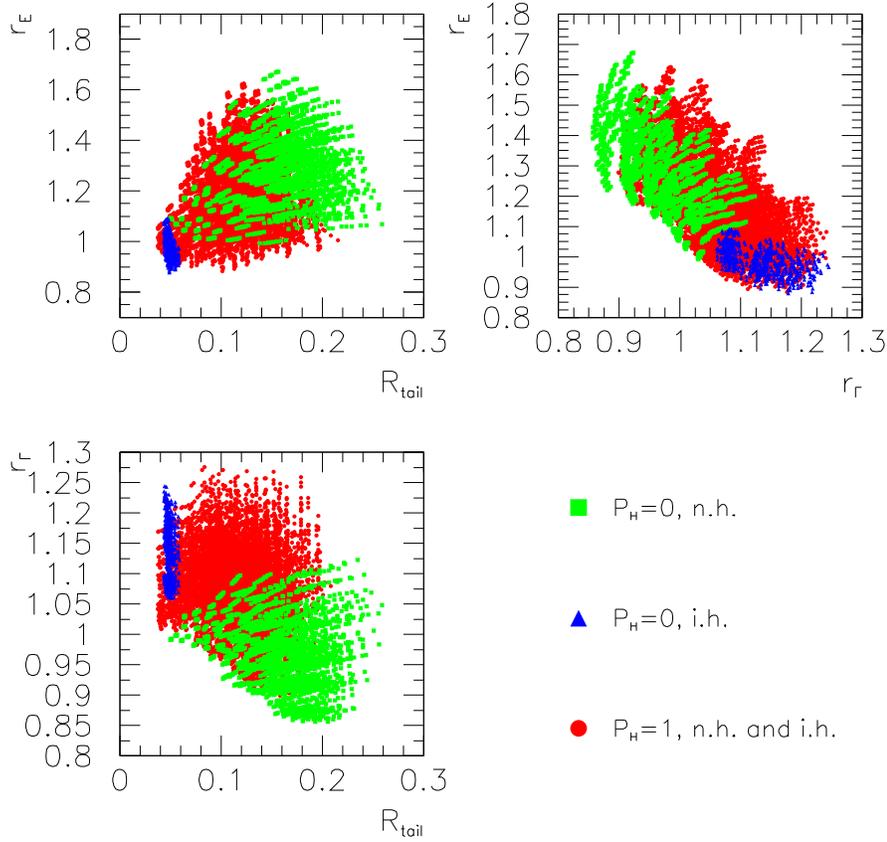}}   
\caption{Scatter plot in the space of the observables $r_E$, $r_{\Gamma}$, $R_{tail}$ for the cases A, B, C discussed in the text. \label{fig:2}}
\end{figure}

Figure \ref{fig:2} shows the regions in the space of the observables
$r_E$, $r_{\Gamma}$, $R_{tail}$ for the cases A, B, C,  obtained by
scanning over the astrophysical parameters in the intervals discussed
in Sec. \ref{sec2}. The values of the oscillation parameters $|\Delta m^2_{32}|$, $\Delta m^2_{21}$, $\theta_{23}$ and $\theta_{12}$ have been taken to coincide with the current best fit points with 10\% error, as expected from near future measurements. To calculate  $R_{tail}$ the cuts $E_L=45$ MeV and $\bar E_L=55$ have been used. 

The results in the figure can be easily interpreted in terms of the
different size of the conversion effects in the different cases
\cite{Lunardini:2003eh}. They show that large regions of the parameter
space exist where only one among the scenarios A, B or C is
possible. Also regions appear where two of these scenarios are
realized.  If these regions are selected by the experiments, the third
possibility will be excluded.








The scenarios in which $0< P_H<1 $ are not shown in 
Fig. \ref{fig:2}. For normal hierarchy and $0<P_H<1$ we expect the
allowed region to be intermediate between the regions found for A and
C. Similarly, for inverted hierarchy and $0<P_H<1$ the region of
possible values of parameters is intermediate between the regions of
cases B and C.  For this reason, the conclusions I derived from
Fig. \ref{fig:2} have essentially an exclusion character and not the
character of establishing one of the scenarios A, B, C.

It is clear that the potential of the method I have discussed here
depends on the statistics and therefore on the distance from the
supernova.  It can be checked \cite{Lunardini:2003eh} that for a
relatively close star ($D\lta 4$ kpc) the error bars are substantially
smaller than the field of points so that the discrimination of
different cases is possible.

\subsection{Analyzing the Earth matter effects}
\label{sub43}

The study of Earth regeneration effects is particularly promising.
The reason is that the information on $\theta_{13}$ and on the mass
hierarchy which can be obtained by this method is largely independent
of astrophysical uncertainties.

This can be understood considering that:

\begin{itemize}

\item The main signature of Earth matter effects -- consisting in
oscillatory modulations of the observed energy spectra (see
Sec. \ref{sec:earth}) -- is unambiguous since it can not be mimicked
by any astrophysical phenomenon. 
Moreover the pattern of oscillation minima and maxima depends only on
$\Delta m^2_{21}$ and $\theta_{12}$, which probably will be
known precisely from solar neutrino experiments and KamLand before the
next galactic supernova event. This will improve the possibility of
identification of Earth matter effects.

\item 
To obtain unambiguous conclusions on $\theta_{13}$ and on the mass
hierarchy it is enough that experiments exclude or establish the Earth
matter effect, without measuring its size precisely, especially if the
effect is probed in both the $\nue$ and $\barnue$ channels.  The fact
that precision is not necessary is very important in this specific
problem, where large astrophysical uncertainties are present.

\end{itemize}

To illustrate the latter point in more detail (see also Table
\ref{tab:obs} for a summary), consider the following scenario. 
\\
\noindent
(i) At
the time of arrival of the neutrino burst, at least one running
detector is shielded by the Earth. 
\\
\noindent
(ii) The shielded detectors record data due to both $\nue$ and
$\barnue$. The two ($\nue$ and $\barnue$) sets of events can be
efficiently separated and for each of them the energy spectrum of the
incoming neutrinos can be reconstructed with good resolution.
\\
\noindent
(iii) The statistics of both the data sets are sufficiently large so
that (oscillatory) spectral distortions as large as  $\sim 10 - 20\%$ can
be established with high statistical significance.  
\\

If these conditions are fulfilled, 
we have four possible experimental results. The corresponding
conclusions on $\theta_{13}$ and on the neutrino mass hierarchy
are discussed below and summarized in Table \ref{tab:obs}. They easily
follow from Eqs. (\ref{fluxesearth}) and (\ref{fluxesearth_ih}). For
simplicity two possibilities are considered in the discussion. The
first is the case in which
substantial differences in the fluxes of different flavors are assumed,
on the basis of (future) precise theoretical predictions. In the second
case astrophysical uncertainties are large and allow equality of
the original fluxes. The generalization of the discussion to
intermediate cases is straightforward.
\\

1. \underline{The Earth effects are established in both $\nue$ and
   $\barnue$ channel.} In this case \emph{unique} conclusions are
   obtained on the oscillation parameters and on the original fluxes
   at the same time.  A first result is the difference of the original
   fluxes in the different flavors: $F^0_e \neq F^0_x$ and $F^0_{\bar
   e} \neq F^0_{\bar x}$. This would be an important test of
calculations of neutrino spectra formation inside the star.  Secondly,
we get that $P_H$ is significantly different from zero, $P_H \sim
1$. This gives the upper bound $\sin^2 \theta_{13} \lta 10^{-6}$ (see
Sec. \ref{sec3}).  The mass hierarchy remains undetermined.  \\

2. \underline{The Earth effect is seen in the $\nue$ channel only},
   and excluded in the $\barnue$ channel. This tells us that $F^0_e
   \neq F^0_x$.  If $F^0_{\bar e} \neq F^0_{\bar x}$ is assumed, again
   the conclusion is unambiguous: the mass hierarchy is inverted and
   $P_H\sim 0$,
   corresponding to $\sin^2 \theta_{13} \gta 10^{-4}$.  In absence of
    {\it a priori} assumptions on the fluxes, we must consider that the
   equality $F^0_{\bar e} \simeq F^0_{\bar x}$ could suppress the
   Earth matter effect on antineutrinos, allowing  other scenarios
   of hierarchy and $\theta_{13}$.  Nevertheless, the case of normal
   hierarchy with $P_H\sim 0$ remains excluded.  \\

3. \underline{The Earth effect is established in the $\barnue$ channel
   only}, while excluded in the $\nue$ channel.  Similarly to the
   previous case, here we conclude that $F^0_{\bar e} \neq F^0_{\bar
   x}$. Assuming that the neutrino original fluxes are different, we
   have that the normal hierarchy is singled out and $P_H\sim 0$
   ($\sin^2 \theta_{13} \gta 10^{-4}$).  The possibility that $F^0_{
   e} \simeq F^0_{ x}$ appears exotic, since both numerical
   calculations and simple physical considerations predict that the
   difference $(F^0_x - F^0_e)$ should be larger than $(F^0_{\bar x} -
   F^0_{\bar e})$.  If the case $F^0_{ e} \simeq F^0_{ x}$ is allowed,
   still the scenario of inverted hierarchy
   with $P_H\sim 0$ is excluded. \\


4. \underline{No Earth effect is seen in both $\nue$ and $\barnue$
   channel.} As can be easily realized,  this result requires that at
   least in one channel ($\nue$ or $\barnue$) oscillations are
   suppressed by equality of original fluxes: $F^0_e \simeq F^0_x$ or
   $F^0_{\bar e} \simeq F^0_{\bar x}$.  
   This would be interesting as a test of predictions of the neutrino
fluxes and energy spectra. No definite conclusions on oscillation
parameters are possible.  \\

\begin{table}
\begin{center}
\begin{tabular}{lcccccc}
\hline 
\hline 
$\nue$ & $\barnue$ & & CONCLUSIONS: &  \\ 
 &  & fluxes  & hierarchy & $P_H$ ($\sin^2\theta_{13}$) \\ 
\hline 
\hline 
yes & yes & different  & undetermined & $P_H\sim 1 $ ($\sin^2\theta_{13}\lta 10^{-6}$).  \\ 
\hline yes & no &  $F^0_e
   \neq F^0_x$ and  $F^0_{\bar e}
   \neq F^0_{\bar x}$&  inverted & $P_H\sim 0 $ ($\sin^2\theta_{13}\gta 10^{-4}$) \\ 
 & & OR: &   & \\
 & & $F^0_e
   \neq F^0_x$ and $F^0_{\bar e} \simeq F^0_{\bar x}$  &  
\emph{exclusion} ~~of ~~normal  & hierarchy~~ with~~ $P_H\sim 0$ \\
\hline no & yes & different fluxes &  normal  &$P_H\sim 0 $ ($\sin^2\theta_{13}\gta 10^{-4}$)   \\ 
\hline no & no & $F^0_e \simeq F^0_x$ or
   $F^0_{\bar e} \simeq F^0_{\bar x}$ & any of the above & any of the above  \\ 
\hline 
\hline
\end{tabular}
\caption{Summary of conclusions that can be drawn from different cases
of observation (``yes'') or exclusion (``no'') of Earth matter effects
in $\nue$ and $\barnue$ channels.}
\label{tab:obs}
\end{center}
\end{table}

The discussion can be generalized to different experimental setups.
For instance, let us consider the situation in which the Earth matter
effect is probed in the $\barnue$ channel only (data on $\nue$ may not
be available or the $\nue$ detector is not shielded by the Earth).  If
the effect is seen, we can exclude the combination of inverted
hierarchy and $P_H \sim 0$.  If it is not, the same combination is
established provided that differences in the original fluxes are
assumed from theory. If the equality of original fluxes is allowed, no
conclusions are possible (see Table \ref{tab:obs}).  If the Earth
matter effects are probed in the $\nue$ channel only, a similar
argument applies, leading to the exclusion or establishment of the
combination $P_H \sim 0$ with normal hierarchy. Again, results depend
on the size of the uncertainties in the original neutrino fluxes.

\subsection{Remarks on the Earth matter effect}
\label{remarks}

Interesting aspects emerge from the discussion in Sec. \ref{sub43}.
\\

One of them is the importance of combining different data sets.
Probing the Earth matter effects in both the $\nue$ and the $\barnue$
channel is \emph{crucial} to disentangle the information on the 1--3
mixing and on the mass hierarchy from astrophysical uncertainties.  If
the Earth regeneration effect is probed in one channel only the
conclusions on the mass hierarchy and $\theta_{13}$, though
potentially strong, have only a ``conditional'' character, since they
depend on assumptions on the original neutrino fluxes.
\\

A second, very important, point is that a negative result is
\emph{stronger} than a positive one. Indeed, if the Earth effect is
seen on both neutrinos and antineutrinos (case 1. of
Sec. \ref{sub43}), the mass hierarchy remains undetermined. In
contrast, in the event of no observation of the Earth matter effect in
$\nue$ channel, together with a positive result on $\barnue$ (case
3.), one obtains both a lower bound on $\theta_{13}$ and establishes
the normal mass hierarchy.  The inverted hierarchy is excluded for
\emph{any} value of $\theta_{13}$.  
Indeed, as shown in Sec. \ref{sec3}, for inverted hierarchy the H
resonance is in the antineutrino channel. In the neutrino channel
nothing prevents the transition, inside the star, of $\nu_e$ to
$\nu_2$, which then should oscillate in the matter of the Earth.  So
one should see the Earth matter effect in the neutrino channel.

Remarkably, the same conclusion
holds even if the negative result for $\nue$ is the only information
available, provided that the inequality of original fluxes can be
safely assumed from theory.  A similar argument is
valid for the case of non observation in the $\barnue$ channel, with
the cautionary remark that $F^0_{\bar e} \simeq F^0_{\bar x}$ could be
realized and lead to weaker conclusions.

It should be also considered that a negative result on the Earth
regeneration effect in one of the channels could be an indication of
the existence of a fourth, sterile, neutrino species.  The
neutrino mass spectrum would have the so called $(3+1)$ form,
characterized by a strong hierarchy between the fourth mass
eigenstate, predominantly sterile in flavor, and the remaining three,
constituted mainly by active flavors \cite{Peres:2000ic}.
\\

In the event that the study of Earth matter effects is not conclusive
-- on the type of neutrino mass spectrum and on the 1--3 mixing -- due
to uncertainties in the original neutrino fluxes, the remaining
ambiguities could be resolved by the combination with other
observations or by the study of specific features of the Earth effect
itself. For instance, the adiabatic region, $\sin^2\theta_{13}\gta
10^{-4}$, could be selected if the Earth matter effect is initially
absent and appears only at late times due to shock-wave effects
\cite{Lunardini:2003eh}. The channel in which this happens would also
determine the mass hierarchy.

Shock-wave effects could also confirm or exclude the presence of a
sterile neutrino, together with other results like, e.g., the absence
of the early peak in the $\nue$ luminosity \cite{Peres:2000ic}
expected from the capture of electrons on protons in the inner regions
of the star (neutronization peak).

\section{Perspectives}
\label{pers}

To conclude, I mention two directions in which progress would be
particularly important for the reconstruction of the neutrino mass
hierarchy and the 1--3 mixing with supernova neutrinos.

On the experimental side, it would be crucial to improve the
experimental sensitivity to electron neutrinos.  While about $10^4$
electron antineutrinos can be detected by SuperKamiokande for a
galactic supernova at $10$ kpc distance (see, e.g.,
\cite{Takahashi:2001dc}), the detection of $\nue$ with existing
detectors (e.g. SNO \cite{waltham}) would give at most few hundreds of
events and would be affected by background processes.  
In this situation, any combined analysis of $\nue$ and $\barnue$ data
would be limited by the large statistical errors affecting the $\nue$
channel.  Therefore, it is clear that to have a new detector with good
energy resolution and capable of recording $\sim 10^3$ or more $\nue$
events would be highly desirable. This experiment should be able to
operate for decades (given the low rate of core collapse of stars in
our galaxy) with no significant interruptions.
Proposals exist to detect electron neutrinos from supernovae using
liquid Argon \cite{Cline:2001pt,bueno} or water enriched with
Gadolinium \cite{beacomvagins}. Other targets like lead, noble gases
or other nuclei have also been considered (see
e.g. \cite{Elliott:2000su,Engel:2002hg,Horowitz:2003cz}).  The
detailed study of the potential of these and other projects for the
reconstruction of the neutrino mass hierarchy and $\theta_{13}$ --
depending on the size of astrophysical uncertainties -- is among the
main goals of research.
  
On the theoretical side, the improvement of predictions of the
neutrino original spectra in the different flavors would be important.
The goal should be to reach consensus between different models and
precision in the numerical calculations.  
The conditions for a
detailed study of supernova neutrinos would be optimal if the
calculation of the neutrino spectra could be done within a model which
naturally predicts the explosion of the star. 

\section*{Acknowledgments}
I would like to thank the organizers and the participants of SUSY 2003
for the stimulating atmosphere I enjoyed there. I am grateful to
A.~Yu.~Smirnov for useful discussions and precious comments on the
manuscript. The research work presented in these proceedings was
supported by the Keck fellowship and the NSF grants PHY-0070928 and
PHY99-07949.

\bibliography{susy}

\end{document}